\documentclass[aps,preprint,groupedaddress,amssymb,amsmath,floatfix]{revtex4-2} %
\usepackage{xcolor}
\usepackage[normalem]{ulem}

\newcommand{\moire}{moir\'e\:}

\newcommand{\MoSe}{MoSe$_{2}$\:}
\newcommand{\SiN}{Si$_{3}$N$_{4}$\:}

\usepackage[colorlinks=true, allcolors=blue]{hyperref}
\usepackage[normalem]{ulem}
\usepackage{graphicx}
\usepackage{todonotes}
\usepackage{acronym}
\graphicspath{ {./figures/} }
\usepackage{amsmath}
\usepackage[LGRgreek]{mathastext}
\usepackage{lineno}
\usepackage[english]{babel}
\usepackage{multibib}
\newcites{methods}{References}
\begin{document}

\title{Topological Fermi-polaron Polaritons with \\ Electrically Invertible Topology}
\author{Xin Xie$^{1,2}$}
\author{Chenxi Liu$^{1,3}$}
\author{Yuze Liu$^{4}$}
\author{Lingxiao Zhou$^{1}$}
\author{Chulwon Lee$^{1}$}
\author{Yuhan Zhang$^{5}$}
\author{Nathanial Lydick$^{1}$}
\author{Kenji Watanabe$^{6}$}
\author{Takashi Taniguchi$^{7}$}
\author{Kai Sun$^{1}$}
\author{Hui Deng$^{1,4*}$}

\affiliation{$^1$Department of Physics, University of Michigan, Ann Arbor, Michigan 48109, United States}
\affiliation{$^2$Michigan Institute for Data $\&$ AI in Society, University of Michigan, Ann Arbor, Michigan 48109, United States}
\affiliation{$^3$Department of Nuclear Engineering, University of Michigan, Ann Arbor, Michigan 48109, United States}
\affiliation{$^4$Department of Electrical Engineering and Computer Science, University of Michigan, Ann Arbor, Michigan 48109, United States}
\affiliation{$^5$Applied Physics Program, University of Michigan, Ann Arbor, Michigan 48109, United States}
\affiliation{$^6$Research Center for Electronic and Optical Materials, National Institute for Materials Science, 1-1 Namiki, Tsukuba, 305-0044, Japan}
\affiliation{$^7$Research Center for Materials Nanoarchitectonics, National Institute for Materials Science,  1-1 Namiki, Tsukuba 305-0044, Japan}
\affiliation{$^*$dengh@umich.edu}

\begin{abstract}
Topological polaritons provide a route to engineer optical Chern bands, but realizing large gaps or efficient electrical tunability remains challenging because time-reversal-symmetry breaking in semiconductor polariton systems is typically limited by weak Zeeman splittings. Here, we demonstrate a topological Bose-Fermi mixture--Fermi-polaron polaritons--where large gaps and electrically invertible topology emerge from the interplay of Fermi-polaron many-body interactions, photonic symmetry, and strong light--matter coupling. Using a gate-controlled monolayer transition-metal dichalcogenide integrated with a symmetry-engineered photonic crystal, we create Fermi-polaron polaritons with topological band structures. The direct coupling between band topology and gate-tunable many-body interactions provides a mechanism for amplified time-reversal-symmetry breaking beyond bare Zeeman splitting and enables electrical inversion of topological gaps, Berry curvature, and Chern numbers. These results establish an electrically reconfigurable platform for interaction-driven topological states and programmable topological photonics.

\end{abstract}
\maketitle

\newpage

\section*{Introduction}
Topological band structures provide a powerful framework for engineering quantum states characterized by global geometric invariants~\cite{vonKlitzing_quantized_1986,Bansil_Colloquium_2016}. Following seminal theoretical works~\cite{Raghu_Analogs_2008,Haldane_Possible_2008}, studies of topological phenomena have been experimentally realized across a broad range of artificial lattices and photonic platforms~\cite{Lu_Topological_2014,Ozawa_Topological_2019}, ranging from atomic gases in optical lattices~\cite{Jotzu_Experimental_2014,Goldman_Light_2014,Goldman_Topological_2016}, electron or hole gases in \moire lattices~\cite{Spanton_Observation_2018,Nuckolls_Strongly_2020,li_Quantum_2021,Cai_Signatures_2023,Xu_Observation_2023,Zeng_Thermodynamic_2023,li_Signatures_2026}, photonic~\cite{Hafezi_Imaging_2013,Lustig_Photonic_2019,Wang_Observation_2009,Rechtsman_Photonic_2013,Jin_Floquet_2025} and polariton platforms~\cite{Klembt_Excitonpolariton_2018,Gianfrate_Measurement_2020,Polimeno_Tuning_2021,Liu_Generation_2020,Li_Experimental_2021,Peng_Topological_2024,Zhao_Subpicosecond_2025,Guddala_Topological_2021,Smirnova_Polaritonic_2024,
Guddala_Topological_2025}. 
Among optical platforms, spin and valley Halls, gapped by breaking spatial symmetries while preserving time-reversal symmetry (TRS), have been widely demonstrated~\cite{Hafezi_Imaging_2013,Liu_Generation_2020,Li_Experimental_2021,Peng_Topological_2024,Zhao_Subpicosecond_2025,Guddala_Topological_2021,Smirnova_Polaritonic_2024,
Guddala_Topological_2025}. They support counter-propagating boundary modes protected against symmetry-preserving perturbations~\cite{Liu_Generation_2020,Li_Experimental_2021,Peng_Topological_2024,Zhao_Subpicosecond_2025}, which require engineered domain boundaries and have inherently finite crosstalk~\cite{Khanikaev_Topological_2024,Arregui_Quantifying_2021,Rosiek_Observation_2023,Saba_Nature_2020}. 
By contrast, TRS-broken Chern bands support unpaired chiral edge modes in the bulk gap, immune from backscattering and robust against interface geometry and local perturbations~\cite{Khanikaev_Topological_2024}. Yet realizing optically accessible Chern bands with large topological gaps or in situ tunability remains a central challenge, 
in part due to limited TRS breaking strength in non-magnetic solids that caps the gap size~\cite{Klembt_Excitonpolariton_2018,Gianfrate_Measurement_2020,Polimeno_Tuning_2021}.

Hybrid platforms that combine two-dimensional semiconductors with photonic crystals (PhCs) have been theoretically proposed to support larger-gap Chern polaritons by combining strong TRS breaking in excitons with symmetry-engineered PhC modes~\cite{He_Polaritonic_2023,Xie_Polariton_2025}. In particular, PhC bands with phase winding around Brillouin zone center can enable direct optical accessibility~\cite{Xie_Polariton_2025}. But their experimental realization has been hampered by the difficulty to achieve sufficient TRS breaking, and electrical control has remained unexplored.

Here we show that Bose-Fermi mixtures -- Fermi-polaron polariton~\cite{Sidler_Fermi_2017,Lyons_Giant_2022,Cotlet_Transport_2019,Chervy_Accelerating_2020,Tan_Interacting_2020,Khestanova_Electrostatic_2024} -- open a pathway for electrically tunable, strong TRS breaking beyond Zeeman splitting.
Using a gate-tunable \MoSe monolayer strongly coupled to symmetry-protected quadratic-touching bands in a photonic crystal, we demonstrate topological Fermi-polaron polaritons with up to 4~meV gap and electrically invertible topology.
Under a magnetic field, the Fermi-polaron many-body interactions, combined with strong exciton-photon coupling, leads to TRS breaking that is an order of magnitude greater than the exciton Zeeman splitting. The resulting gap opening at original touching point $\Gamma$ reaches up to $\sim$ 4~meV under a 7~T field, while previous bosonic polariton systems relying primarily on Zeeman splitting have been inherently limited to about 0.1--0.5~meV~\cite{Klembt_Excitonpolariton_2018,Gianfrate_Measurement_2020,Polimeno_Tuning_2021}. Furthermore, in this many-body interacting system, topology of the bands and the density of the electron or hole gas become directly coupled. By adjusting the gate voltage, we demonstrate the unique ability to invert the topological charge (Chern number) and sign of the Berry curvature on demand, without flipping external magnetic fields, as well as to continuously tune the topological gap sizes and distributions of quantum geometries. 
By combining PhC design with many-body interactions and electrical tunability in a Bose-Fermi mixture, this platform offers a route toward programmable topological photonic circuitry and robust polaritonic routing functionalities~\cite{Cheng_Robust_2016}.

\section*{Principles}
Our device consists of a gated \MoSe\ ML, encapsulated by hexagonal boron nitride (hBN) and placed on a 2D PhC (Fig.~\ref{f1}a). Details of fabrication are provided in Methods. 

Using the device, distinct from previous approaches~\cite{Wang_Observation_2009,Rechtsman_Photonic_2013,Klembt_Excitonpolariton_2018,Lustig_Photonic_2019,Gianfrate_Measurement_2020,Polimeno_Tuning_2021,Jin_Floquet_2025}, we create topological bands by combining three key ingredients: a PhC with designable spatial geometry that introduces phase winding around the optically accessible $\Gamma$-point in Fourier space (Fig.~\ref{f1}b), a valley-polarized carrier reservoir, referred to here as a Fermi sea, controlled by electrical and magnetic fields to introduce much stronger TRS breaking than that of single-particle systems (Fig.~\ref{f1}c), and strong light-matter coupling that combines the spatial- and time-symmetry properties to create topological band structures (Fig.~\ref{f1}d). Below we discuss each in more details. 

First, as illustrated in Fig.~\ref{f1}b, to produce geometric phase for topological band engineering, we use quadratic-touching bands with symmetry-protected degeneracy at the $\Gamma$ point, the center of both the light cone and the Brillouin zone~\cite{Xie_Polariton_2025}. This allows us to combine direct optical access to high-symmetry points and large bandwidths, lifting limitations on gap size imposed by the Bloch bandwidth. 
In PhCs with three-fold, four-fold, and six-fold rotational symmetries, the degeneracy at $\Gamma$ is jointly protected by spatial symmetry and TRS. 
Away from $\Gamma$, reduced symmetry leads to hybridization of the two eigenmodes with a quadratic winding coupling $\nu k^2 e^{i2\phi_k}$, manifested as quadratic splitting and in-plane rotation of the eigenstates (see Methods and Sec.~I in Supplementary Information for further details). When the symmetry-protected degeneracy at $\Gamma$ is lifted by breaking TRS, the resulting bands acquire non-trivial Chern numbers and Berry curvature~\cite{Xie_Polariton_2025}. 

Second, we control TRS via Fermi polaron modes in a gated \MoSe{} ML~\cite{Efimkin_Manybody_2017,Back_Giant_2017}, as illustrated in Fig.~\ref{f1}c. 
In an undoped \MoSe{} ML (middle panel), as a result of broken spatial inversion symmetry and preserved TRS of the crystal, doubly degenerate excitons are formed by electrons and holes of opposite spins, at $K$ and $K'$ valleys, coupled to light of opposite circular polarizations $\sigma^\pm$, respectively, known as spin-valley locking~\cite{Xiao_Coupled_2012,Schaibley_Valleytronics_2016}. A positive magnetic field introduces different Zeeman splittings to the conduction and valance bands, as well as an exciton valley splitting, but only 2-3~meV even under a magnetic field of 10~T~\cite{MacNeill_Breaking_2015}. 

When a Fermi sea is introduced via gating, however, a much stronger effective TRS breaking can be achieved (Fig.~\ref{f1}c, left and right panels). Due to the electron and hole valley splittings, doped carriers fill the lower-energy valley first, while leaving the opposite valley undoped, creating a valley- and spin-polarized Fermi sea.
The Fermi sea interact with excitons in the same and opposite valleys via dominantly Pauli exclusion and Coulomb effects, respectively, forming repulsive (RP) and attractive polarons (AP). These quasiparticles exhibit a polaron splitting $|\Delta E_{P}|$ reaching tens of meV while maintaining appreciable oscillator strength, enabling the formation of Fermi polaron–polaritons and interaction-driven optical and transport effects~\cite{Sidler_Fermi_2017,Lyons_Giant_2022,Cotlet_Transport_2019,Chervy_Accelerating_2020,Tan_Interacting_2020,Khestanova_Electrostatic_2024}. The AP, as well as RP, has a TR-partner in the opposite valley, but with an oscillator strength that can differ by orders of magnitude.

Third, the valley contrast in the Fermi-polaron oscillator strengths can be transcribed to a large TRS-breaking gap via the third key ingredient: strong light-matter coupling, which also provides the mechanism to integrate the TRS breaking with phase winding in PhC. 
When the gated-\MoSe\ ML is integrated on the PhC, 
the polariton bands inherit the symmetry-protected $\Gamma$-point degeneracy and phase winding of the photonic bands. The valley-contrasting AP/RP oscillator strengths then induce valley-selective coupling, breaking TRS and producing helicity-dependent splitting of the Fermi-polaron polariton branches. This splitting acts as an effective TRS-breaking mass term that gaps the quadratic touching and generates nontrivial Berry curvature (see Methods and Sec.~II of the Supplementary Information for details). Figure~\ref{f1}d illustrates a simplified limiting case of topological Fermi-polaron polaritons in which the AP and RP are fully polarized in opposite valleys, with their time-reversed partners suppressed. Consequently, the strength of TRS breaking is no longer limited by the valley splitting, but instead by the interaction energies among the Fermi sea, the exciton, and the PhC photon.

Notably, hole and electron doping generate opposite effective TRS-breaking mass terms, reflected in the Bloch-sphere trajectories in Fig.~\ref{f1}d. The pseudospin starts from opposite poles, and evolves toward the equator as the momentum moves away from $\Gamma$ and winds twice with the azimuthal momentum angle $\phi_k$, yielding topological bands with gate-invertible Chern numbers $\pm 1$. 
The gap size can be controlled by the total TRS breaking strength, which depends on the doping density and is tunable by the gate voltage. Therefore, strong polaron-photon coupling provides the mechanism to transfer gate-controlled oscillator-strength imbalance of valley polarons to strong TRS breaking of geometry-engineered PhCs beyond the bare Zeeman splitting, enabling large topological gaps and electrical -- instead of magnetic -- tuning and inversion of topology.

To experimentally establish the electrically tunable topological Fermi-polaron polaritons, we separately characterize the properties of the bare PhC, the bare gated \MoSe{}, and integrated \MoSe-PhC system, all in the same device, where the gated \MoSe{} covers partially both the PhC and the unprocessed \SiN slab. Their eigenmodes are measured using circularly polarized, momentum-resolved magneto-optical reflection contrast (RC) spectroscopy (see Methods and the Supplementary Information Fig. S2 for details). 

\section*{Quadratic-touching PhC Photon}
As illustrated in Fig.~\ref{f2}a, we use a PhC with sixfold rotational symmetry, consisting of a honeycomb lattice of identical inverted triangular air holes patterned into a \SiN\ slab suspended over a Si substrate. Using finite-difference time-domain (FDTD) simulations, we design it to support a pair of quadratic bands with a symmetry-protected degeneracy at $\Gamma$, with energies near that of the \MoSe{} exciton (Fig.~\ref{f2}b). At $\Gamma$, the modes are dark because of symmetry-forbidden radiation. Away from $\Gamma$, the two branches exhibit orthogonal far-field linear polarizations, with polarization axes at $4\phi_k$ and $4\phi_k+\pi/2$, where $\phi_k$ is the azimuthal momentum angle. Figure~\ref{f2}c shows the representative case of $\phi_k=45^\circ$, where the measured polarization-resolved dispersion agrees with the simulation despite linewidth broadening from fabrication imperfections. 

Different from conventional TE/TM modes, the bands arise from a symmetry-defined, time-reversal-related $E_2$ doublet in the sixfold-rotation-symmetric PhC. It is verified by measured far-field polarization vortex as well as the simulated near-field mode profiles shown in Supplementary Information Figs.~S1 and S2. Together with the winding mechanism discussed above, this $E_2$ modal basis provides the parent photonic structure that becomes topological when gapped by TRS breaking. Further details are provided in Sec.~I of the Supplementary Information.

\section*{Fermi polaron}
Next we characterize the polaron modes of the bare MoSe$_2$ ML, on unprocessed \SiN slab without PhCs (Figs.~\ref{f2}d-i). Figures~\ref{f2}(d,e) show circularly polarized RC spectra versus gate voltage $V_{\mathrm{G}}$ under a +7~T magnetic field. Near charge neutrality ($-1.7$~V$<V_{\mathrm{G}}<1.6$~V), the response is dominated by neutral excitons, with a weak AP or localized trion contribution due to inhomogeneity and impurity in the sample. There is a small exciton valley splitting of about 1.6~meV, barely discernible in the spectra at $V_{\mathrm{G}}=0$~V (Fig.~\ref{f2}d) or in the voltage-energy map (Fig.~\ref{f2}e).

Increasing $|V_g|$, APs are formed between the $K$ ($K'$) valley Fermi sea and $K'$ ($K$) valley exciton under hole (electron) doping; they rapidly gain oscillator strength, while APs in the opposite valley remain negligible. This reflects that the Fermi level is in between the split valence (conduction) bands, the $K$ ($K'$) valley hole (electron) density increases, while the $K'$ ($K$) valley is largely undoped~\cite{Efimkin_Manybody_2017,Back_Giant_2017,Wu_Negative_2022}.
As a result, APs acquire a strong valley contrast, with opposite dominant helicities between hole- and electron-doped regimes. RPs exhibit the opposite valley-polarization, but with a smaller oscillator-strength contrast. Figure~\ref{f2}d shows two representative gate voltages, $V_G=-2.1$~V and $1.6$~V, where the AP valley contrast is most pronounced and clearly goes beyond the bare Zeeman response at $V_G=0$~V.
The strong valley contrast persists even at 5~T (Supplementary Information Fig. S6); but it disappears at 0~T (Supplementary Information Fig. S5), where both valleys are equally filled. At larger $|V_G|$, both valleys become increasingly populated, reducing the valley contrast and weakening the effective TRS breaking, along with linewidth broadening and background variation.

The experimental spectra are well reproduced by transfer-matrix-method (TMM) fits (Fig.~\ref{f2}e), from which we obtain the oscillator strengths and resonance energies of the AP and RP in each valley (see Methods). Possible Landau-level effects~\cite{Efimkin_ExcitonPolarons_2018,Smolenski_Shubnikov_2019} are unresolved in the RC spectra; any associated weak renormalization is included phenomenologically in the extracted polaron parameters. Consistent with the spectra in Figs.~\ref{f2}(d,e), the oscillator strengths $f$ shift from excitons or RPs to mainly one of the APs at moderate $V_G$ (Fig.~\ref{f2}f).
The resulting valley contrasts manifest as strong magneto-optical circular dichroism (MOCD) of the APs, defined as $(f_{\sigma^+} - f_{\sigma^-})/(f_{\sigma^+} + f_{\sigma^-})$, reaching near negative unity at $V_G\sim -2.1$~V and a maximal positive value at $V_G\sim 1.6$~V (Fig.~\ref{f2}h). The RPs exhibit the opposite MOCD trend, but with a smaller magnitude. The quantitative differences between hole and electron doping reflect different $g$ factors for the valence and conduction bands, which lead to a stronger valley imbalance in the hole-doped regime than in the electron-doped regime.
At the same time, both RPs and APs experience a larger blueshift in the doped valley than in the undoped valley due to Pauli exclusion, which adds to the Zeeman splitting between RPs and APs (Figs.~\ref{f2}(g,i)). 
These results confirm gate-control of amplified Fermi-polaron splitting and oscillator strength contrasts under a magnetic field, which provide the microscopic ingredients for the effective TRS breaking discussed below.

\section*{Topological Fermi-polaron polariton}
After characterizing the PhC and \MoSe{}, we now measure the \MoSe{}-PhC coupled system, which combines the valley-contrasting properties of \MoSe{} and geometric phase of the PhC through strong light-matter coupling. 
Via momentum-resolved derivative RC spectra of circularly polarized light, we directly measure the polariton dispersions (Fig.~\ref{f3}a) and reflective magnetic circular dichroism (Fig.~\ref{f3}c) at two representative gate voltages $V_{\mathrm{G}}=-2.1$~V and $1.6$~V with the strongest and opposite MOCD of AP under hole and electron doping, respectively.
The measured dispersions are then compared with theoretical predictions from the coupled polaron–PhC Hamiltonian model (white dashed curves); and the full spectra are compared with rigorous coupled-wave analysis (RCWA) simulations (Fig.~\ref{f3}b). The final polariton structure, which goes beyond the simplified schematic in Fig.~\ref{f1}, is determined by the gate-dependent AP/RP resonance energies, oscillator strengths, and detunings from the photonic bands in the actual device.
Both calculations use parameters extracted from the independent bare PhC and polaron measurements discussed in Fig.~\ref{f2} (RCWA simulations and Effective Hamiltonian model in Methods). 
All four AP/RP resonances are included, except that the $\sigma^+$ AP at $V_{\mathrm{G}}=-2.1$~V is omitted because of its near-zero fitted oscillator strength.

In Figs.~\ref{f3}(a,b), for each $V_{\mathrm{G}}$, the spectra for $\sigma^+$ and $\sigma^-$ polarizations are plotted side-by-side. These spectra represent the radiative polarization projections of the same complete polariton eigenmodes, which contain contributions from the photonic modes and all optically active polaron resonances. The white dashed curves highlight the branches with stronger optical visibility in each detection channel; spectra overlaid with all six calculated branches are shown in Supplementary Information Fig.~S7. They together show three pairs of bright polariton bands -- upper, middle, and lower polaritons -- all with characteristic polariton dispersions and anti-crossing. The calculated bands capture the main polariton branch structure, although small offsets between the calculated eigenenergies and derivative-RC extrema are expected, especially at larger momenta, due to linewidth broadening, asymmetric line shapes, spectral overlap, and momentum-dependent background. The polariton bands near $\Gamma$ become dark as their coupling to free space modes is symmetry-forbidden, while the weak features around 0 and $-30$~meV originate from weakly coupled RP and AP absorption in the monolayer optical response.

Crucially, the polariton bands, which touch at $\Gamma$ point at B=0~T, split, and acquire opposite circular polarization near $\Gamma$, most clearly for the lower-polariton pair. In stark contrast, at B=0~T, absent of TRS breaking, $\sigma^\pm$ spectra are essentially identical, the three polariton pairs remain degenerate at $\Gamma$ and predominantly linearly polarized away from $\Gamma$ at all doping densities (Supplementary Information Figs. S3 and S5). This field-induced helicity-dependent splitting is consistent with a TRS-breaking-induced mass term. Further polarization-resolved analysis shows a magnetic-field-induced circular-to-linear polarization evolution from $\Gamma$ to larger momenta, together with an $S_3$-based proxy that qualitatively follows the calculated ring-like Berry-curvature distribution (Supplementary Information Fig.~S4 and Secs.~III, IV). The Berry curvature vanishes at $k=0$, a characteristic feature of the gapped quadratic-touching bands considered here and distinct from gap opening at Dirac points~\cite{Xie_Polariton_2025}. For the middle-polariton branches, a polarization-dependent spectral difference is observed and agrees qualitatively with the model, but the broader linewidth prevents a reliable quantitative extraction of the gap.

Notably, the helicity patterns invert between hole and electron doping (Figs.~\ref{f3}a--c). Figure~\ref{f3}c shows the corresponding sign reversal of the RMCD signal, most clearly for the lower polaritons. The moderate RMCD magnitude reflects the radiative projection of the polariton eigenstates and linewidth-related spectral overlap, rather than a direct measurement of the intrinsic circular polarization of each branch. This chirality inversion is accompanied by a corresponding reversal of the calculated Berry curvature, as shown in Fig.~\ref{f3}d, where the polariton band dispersions are overlaid with Berry curvature as the color scale. The chiral response remains robust at 5~T (Fig.~S6), disappears at 0~T (Fig.~S5), and reverses under magnetic-field reversal (Fig.~S8).

Here, the sign of Chern number and Berry curvature are no longer dictated only by the direction of the external magnetic field; instead, they depend on the fermionic charge, controlled by the gate voltage. Furthermore, this many-body mechanism produces effective TRS breaking beyond the bare Zeeman splitting alone, enabling a sizable topological gap.

Lastly, we further quantify the effective TRS-breaking-induced mass term by extracting the gap opened at $\Gamma$ versus gate voltage which depends on the combination of gate-dependent AP/RP valley splittings, oscillator-strength valley contrast, and detunings from the photonic bands. Figure~\ref{f4}a shows line cuts of derivative RC spectra of the $\sigma^+$ (red) and $\sigma^-$ (blue) lower polaritons near $k \approx 0$ at three representative gate voltages, with the resonance energy marked by vertical dashed lines. The $\sigma^\pm$ spectra are nearly degenerate at charge neutrality. Upon doping, however, a clear energy splitting between them $\Delta E_{LP}=E_{LP}^{\sigma^+}-E_{LP}^{\sigma^-}$ emerges, and the order of the branches reverses between electron and hole doping. 

Limited by the polariton linewidth of 8–15 meV, we extract the gate-voltage dependence of the $\sigma^\pm$ lower polariton energies by polarization-resolved Lorentzian fits (Fig.~S9 in Supplementary Information), as shown in Fig.~\ref{f4}b. The results show gate tuning of both the magnitude and the sign of the polariton gap, which agree well with theoretical expectations (solid lines) and FDTD simulations (dashed lines). As summarized in Fig.~\ref{f4}c, the gap grows from nearly zero at charge neutrality, reverses sign with gate voltage and reaches a maximum of $\sim$4~meV, more than an order of magnitude larger than in previously reported exciton-polariton platforms~\cite{Klembt_Excitonpolariton_2018,Gianfrate_Measurement_2020,Polimeno_Tuning_2021}. 
Above $V_G\sim1.8$~V, the gap decreases rather than saturates because the valley contrast weakens at higher electron doping; the broader plateau on the hole-doped side is consistent with the stronger hole-side valley imbalance discussed above.
The corresponding Berry-curvature evolution in Supplementary Information Fig.~S10 shows a sign reversal consistent with the gap inversion. As the $\Gamma$-centered gap increases, the Berry curvature broadens in momentum space and its peak amplitude decreases.
These results demonstrate electrically invertible and tunable topological gaps and voltage-driven phase transitions in Fermi-polaron polariton Chern bands.  

\section*{Conclusion}
In short, we demonstrate an electrically-tunable and invertible topological Fermi-polaron polariton on a chip.
In this system, Fermi-polaron many-body interactions and strong light-matter coupling work together to achieve TRS-breaking-induced gaps that are more than an order of magnitude greater than those attainable in existing semiconductor polariton systems that rely primarily on Zeeman effect~\cite{Klembt_Excitonpolariton_2018,Gianfrate_Measurement_2020,Polimeno_Tuning_2021}. 
It allows fast, scalable, and locally addressable electrical tuning and inversion of chirality, gap and quantum-geometry of the hybrid bands without reversing the external magnetic field. Future devices with engineered domain-wall geometries may enable direct measurement of edge-state transport and integer Chern numbers~\cite{Vakulenko_NearField_2021}. In particular, spatially nonuniform gating could create interfaces between regions with opposite electrically controlled topological mass, supporting electrically controllable chiral edge transport. The topological Fermi-polaron polariton system forms a designable on-chip platform for controlled studies of Fermi-sea-mediated topological order, opening a pathway toward electrically programmable topological photonic devices and quantum information processing~\cite{Cheng_Robust_2016}.
%

\section*{Methods}
\textbf{Fabrication of the device.} The PhC structure was fabricated from a 140-nm-thick \SiN layer deposited on a Si substrate by low-pressure chemical vapor deposition. The electrode pattern was defined in SPR 220 photoresist using optical lithography (Heidelberg µPG 501 Mask Maker), followed by gold deposition with an Angstrom Engineering Evovac electron-beam evaporator. The 2D PhC patterns were then written into ZEP 520A e-beam resist using electron-beam lithography (JEOL JBX-6300FS). Subsequently, inductively coupled plasma reactive ion etching (ICP/RIE, STS APS DGRIE) was employed to transfer the pattern into the \SiN layer, then XeF$_2$ gas etching removed the underlying Si substrate to form suspended membranes. Finally, the residual resist was removed by oxygen plasma etching.  

\MoSe ML was mechanically exfoliated from bulk crystals (HQ Graphene) using adhesive tape, together with graphene and hBN flakes. These 2D materials were then sequentially stacked onto the PhC using a conventional dry-transfer method. The fabrication process and the optical images are shown in Supplementary Information Fig. S11.

\textbf{Magneto-optical reflection contrast spectroscopy.} Magneto-optical spectroscopy was performed at 4~K by mounting the sample in an Attodry 1000 system equipped with a superconducting magnet providing an out-of-plane magnetic field (Supplementary Information Fig. S12). RC measurements were carried out by directing broadband white light from lamp or super-continuum (SC) laser with either $\sigma^+$ or $\sigma^-$ circular polarization onto the sample, and collecting the reflected signal in both real and Fourier space from the MoSe$_2$ region ($R$) and the adjacent bare \SiN substrate ($R_{\mathrm{sub}}$). The RC was calculated by $R/R_{\mathrm{sub}}$. The derivative RC was calculated by $d(R/R_{\mathrm{sub}})/dE$. Momentum-resolved circular dichroism was then calculated as $(RC_{\sigma^+} - RC_{\sigma^-})/(RC_{\sigma^+} + RC_{\sigma^-})$.

\textbf{Transfer Matrix Method (TMM) fitting.} We obtain the oscillator strengths and resonance energies of exciton and Fermi polarons in MoSe$_2$ by fitting the measured circularly polarized RC spectra using the TMM. The heterostructure stack consists of hBN/MoSe$_2$/hBN/Si$_3$N$_4$ layers on a Si substrate, with layer thicknesses of 7.5 nm/0.7 nm/2.4 nm/141 nm, respectively, measured by atomic force microscope (AFM) for the hBN and \MoSe layers and by a Woollam M-2000 ellipsometer for the SiN layer. 
For the dielectric layers, we use a constant refractive index in the wavelength range of interest with n=2.2 for hBN, n=2.0 for \SiN, and n=3.5 for Si. 
For the active medium of MoSe$_2$ ML, we use a dispersive dielectric permittivity function $\epsilon(\omega)$ that consists of a constant background permittivity $\epsilon_b$ and dispersive Lorentz terms due to exciton or Fermi polaron resonances: 
\begin{equation}
\epsilon(\omega) = \epsilon_b + \sum_{j} \frac{f_j}{\omega_j^2 - \omega^2 - i\gamma_j \omega},
\end{equation}
where $\omega_j$, $f_j$, and $\gamma_j$ are the resonance frequency, oscillator strength, and linewidth of the $j^{\mathrm{th}}$ resonance, respectively.  
RP, AP for the two helicities ($\sigma^+$ and $\sigma^-$) are modeled with different Lorentz oscillators, four in total, to capture the evolution from exciton to polarons as well as the valley-dependent oscillator strengths and valley splitting. 

In the fitting procedure, the measured spectrum at each gate voltage is fitted with TMM calculations using Matlab’s nonlinear curve fitting function, and fitting parameters includes the polaron resonance energy, oscillator strength, and linewidth. The gate-dependent oscillator strengths, resonance energies and linewidths obtained from the fit are used as inputs to the effective Hamiltonian model and RCWA simulations described below.

\textbf{RCWA simulations.} Rigorous coupled-wave analysis (RCWA) was used to simulate the momentum-resolved RC ($R/R_{sub}$) of bare photonic mode and derivative RC  ($d(R/R_{sub})/dE$) of the \MoSe-PhC heterostructure, including the reflection from both PhC ($R$) and the bare \SiN substrate($R_{sub}$). The complex reflection matrix elements $r_{ss}$, $r_{sp}$, $r_{pp}$, and $r_{ps}$ were calculated and transformed from the linear polarization basis into the circular polarization basis to obtain $RC_{\sigma^+}$ and $RC_{\sigma^-}$. The MoSe$_2$ ML was modeled as an anisotropic dielectric layer whose in-plane permittivity tensor incorporates four Lorentz oscillators with corresponding polarizations, corresponding to the AP and RP modes in the $K$ and $K'$ valleys. The parameters of these oscillators were taken from the TMM fits to the experimental RC spectra, ensuring consistency between the fitting and simulation frameworks. This formulation captures both the valley-dependent Zeeman splitting and the oscillator-strength imbalance between two helicities, enabling direct comparison of simulated and measured RC spectra.

\textbf{Effective Hamiltonian model.} 
The symmetry-protected quadratic-touching bands of the PhC can be represented in the basis of the two degenerate eigenmodes of the six-fold rotation operator: $\{|P^+\rangle, |P^-\rangle\}$. The effective Hamiltonian for the PhC takes the form:
\begin{equation}
H_{\mathrm{PhC}}(\mathbf{k})=
\begin{bmatrix}
\omega_{\mathbf{k}}^{P} & \nu k^2 e^{-i2\phi_{\mathbf{k}}} \\
\nu k^2 e^{i2\phi_{\mathbf{k}}} & \omega_{\mathbf{k}}^{P}
\end{bmatrix},
\end{equation}
where $\omega_{\mathbf{k}}^{P}$ is the dispersion of $|P^\pm\rangle$, $\nu$ is a coupling coefficient, $\mathbf{k}$ is the in-plane wave vector, and $\phi_{\mathbf{k}}$ is the azimuthal angle of $\mathbf{k}$ that counts for the phase winding in coupling. This Hamiltonian gives rise to two quadratic bands, with a degeneracy at $\Gamma$ as required by the combined $C_6$ symmetry and TRS.

Due to the large mismatch in effective masses of the PhC modes and Fermi polarons, polarons in MoSe$_2$ were treated as flat, highly degenerate bands in the photonic Brillouin zone and collectively couple with each PhC mode~\cite{Tavis_Exact_1968}. With two valleys, $K$ or $K'$, there are a total of four APs and RPs coupled with each PhC mode. The effective Hamiltonian $H_{\mathbf{k}}$ for the coupled system can be expressed in a ten-component basis:
\[
\Phi = (|P^+\rangle, |K_{RP}^+\rangle, |K_{RP}'^+\rangle,|K_{AP}^+\rangle, |K_{AP}'^+\rangle, |P^-\rangle, |K_{RP}^-\rangle, |K_{RP}'^-\rangle,|K_{AP}^-\rangle, |K_{AP}'^-\rangle),
\]
where $|K_{RP/AP}^\pm\rangle$, $|K_{RP/AP}'^\pm\rangle$ represent collective RP and AP modes in the $K$ and $K'$ valleys coupled to $|P^\pm\rangle$. In this basis, we have:
\begin{equation}
H_{\mathbf{k}}=
\begin{bmatrix}
\omega_{\mathbf{k}}^{P} & \alpha_K^{RP} & \beta_{K'}^{RP}& \alpha_K^{AP} & \beta_{K'}^{AP} & \nu k^2 e^{-i2\phi_{\mathbf{k}}} & 0 & 0  & 0 & 0\\
\alpha_K^{RP*} & \omega_{K}^{RP} & 0 & 0 & 0 & 0 & 0 & 0 & 0& 0\\
\beta_{K'}^{RP*} & 0 & \omega_{K'}^{RP} & 0 & 0 & 0 & 0 & 0 & 0& 0\\
\alpha_K^{AP*} & 0 & 0& \omega_{K}^{AP} & 0 & 0 & 0 & 0 & 0 & 0\\
\beta_{K'}^{AP*}  & 0 & 0 & 0 & \omega_{K'}^{AP} & 0 & 0 & 0 & 0& 0\\
\nu k^2 e^{i2\phi_{\mathbf{k}}} & 0 & 0 & 0& 0 & \omega_{\mathbf{k}}^{P} & \beta_K^{RP*} & \alpha_{K'}^{RP*} & \beta_K^{AP*} & \alpha_{K'}^{AP*}\\
0 & 0 & 0 &  0 & 0 &\beta_K^{RP} & \omega_{K}^{RP} & 0 & 0 & 0 \\
0 & 0 & 0 & 0 & 0 &\alpha_{K'}^{RP} & 0 & \omega_{K'}^{RP}&0 & 0 \\
0 & 0 & 0 &  0 & 0 &\beta_K^{AP}  & 0 & 0& \omega_{K}^{AP} & 0  \\
0 & 0 & 0 & 0 & 0 &\alpha_{K'}^{AP}&0 & 0  & 0 & \omega_{K'}^{AP}\\
\end{bmatrix},
\label{eq:Hk_quadratic}
\end{equation}
where $\omega_{K}^{RP/AP}$ and $\omega_{K'}^{RP/AP}$ are the RP and AP energies in the two valleys, and $\alpha_{K/K'}^{RP/AP}$, $\beta_{K/K'}^{RP/AP}$ are the collective light–matter coupling strengths. These are related to the oscillator strengths $f_{K/K'}^{RP/AP}$ via $\alpha_{K/K'}^{RP/AP} = \alpha \sqrt{f_{K/K'}^{RP/AP}}$ and $\beta_{K/K'}^{RP/AP} = \beta \sqrt{f_{K/K'}^{RP/AP}}$. The ratio $\alpha/\beta$ is determined by the decomposition of the PhC mode into the circular basis, with the two PhC modes related by TRS, yielding conjugate couplings for the partner mode. $\alpha$ and $\beta$ were calculated from FDTD simulations of the photonic modes. The oscillator strengths $f_{K/K'}^{RP/AP}$ and energies $\omega_{K/K'}^{RP/AP}$ of polarons were determined from TMM fits to the RC spectra. Polariton dispersions were obtained from the eigenvalues of the effective Hamiltonian, while the Berry curvature and quantum metric were calculated from its eigenstates. Specifically, the Berry curvature and Chern numbers were evaluated using a gauge-invariant Wilson-loop formulation on a discretized momentum-space mesh, following the lattice Berry-curvature method~\cite{Fukui_Chern_2005,Yu_Equivalent_2011}.\\

\noindent\textbf{Data availability}

\noindent Data will be deposited in the University of Michigan institutional repository, Deep Blue, upon publication. \\

\noindent\textbf{Acknowledgment}

\noindent X.X, Y.L., L.Z., C.L. and H.D. acknowledge support from the Army Research Office under Awards W911NF-25-1-0055, the Office of Naval Research under Awards N00014-21-1-2770, the National Science Foundation under Awards EECS 2529062, and the Gordon and Betty Moore Foundation under Grant GBMF10694. K.S. acknowledge support from the Office of Naval Research under Awards N00014-21-1-2770 and the Gordon and Betty Moore Foundation under Grant GBMF10694. X.X. acknowledges the support by Schmidt Sciences, LLC. K.W. and T.T. acknowledge support from the JSPS KAKENHI (Grant Numbers 21H05233 and 23H02052) , the CREST (JPMJCR24A5), JST and World Premier International Research Center Initiative (WPI), MEXT, Japan. X.X thanks Ruofan Hao, Yudong Wu, Yixuan Chen at University of Michigan for assistance with exfoliation and preparation of the 2D materials. \\

\noindent\textbf{Author Contributions}

\noindent X.X. and H.D. conceived and designed the research. X.X. performed the modeling, simulations, device fabrication, measurements and data analysis. C.Liu prepared the layered materials with input from Y.Z. and assisted with 2D material transfer. C.Lee grew the \SiN wafer. K.W. and T.T. provided the hBN bulk crystals. C.Liu, Y.L., L.Z., and N.L. assisted with experiment setup and measurement. X.X., K.S., and H.D. analyzed the results. X.X. and H.D. wrote the manuscript with contributions from all authors.\\

\noindent\textbf{Competing interests} 

\noindent The authors declare that they have no competing financial interests.

\clearpage
\setcounter{figure}{0}

\begin{figure}[h]
\centering
\includegraphics[width=0.85\linewidth]{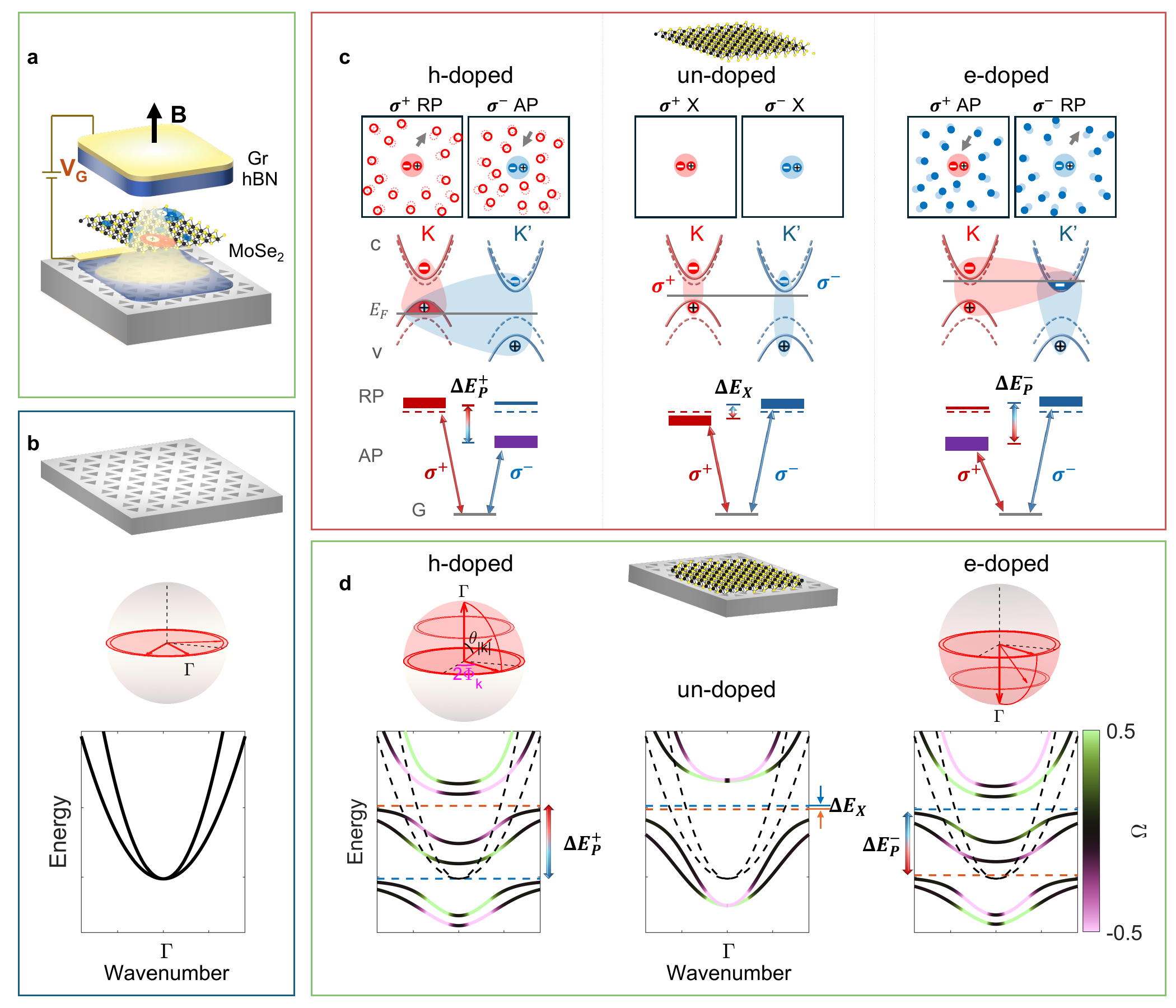}
\caption{\small \textbf{Gate-tunable topological Fermi-polaron polariton in a PhC–\MoSe{} device.}
\textbf{a}, The device under a positive magnetic field. An hBN-encapsulated \MoSe{} ML hosts Fermi polarons that are controlled by a graphene (Gr) top gate and strongly couple with symmetry-engineered PhC modes.
\textbf{b}, A PhC slab (top) and its bands with symmetry-protected quadratic touching at $\Gamma$ point (bottom). The PhC eigenstates rotate around the equator when moving across the Brillouin zone (middle), corresponding to bands with zero Berry curvature.
\textbf{c}, Spin selectivity of RP and AP in \MoSe\ ML with hole doping (left column), no doping (middle column) and electron doping (right column) under a positive magnetic field. Top to bottom: ML atomic lattice; spin-polarized RP and AP or excitons in opposite valleys in real space; electronic band structures; exciton and polaron energy levels. Red (blue) represents $K$ ($K'$) valley.   
\textbf{d}, Gate-tunable topological polaron-polaritons for hole-doped (left), undoped (middle), and electron-doped (right) regimes. Shown are the device schematic, evolution of upper polariton eigenstates exhibiting spin-dependent double wrapping of half Bloch sphere (Chern number $\pm1$), and polariton band dispersions with Berry curvature $\Omega$ represented by color. Black dashed curves indicate uncoupled photonic bands; red and blue dashed lines mark the $\sigma^+$ and $\sigma^-$ polaron branches.}\label{f1}
\end{figure}

\newpage

\begin{figure}[h]
\centering
\includegraphics[width=0.75\linewidth]{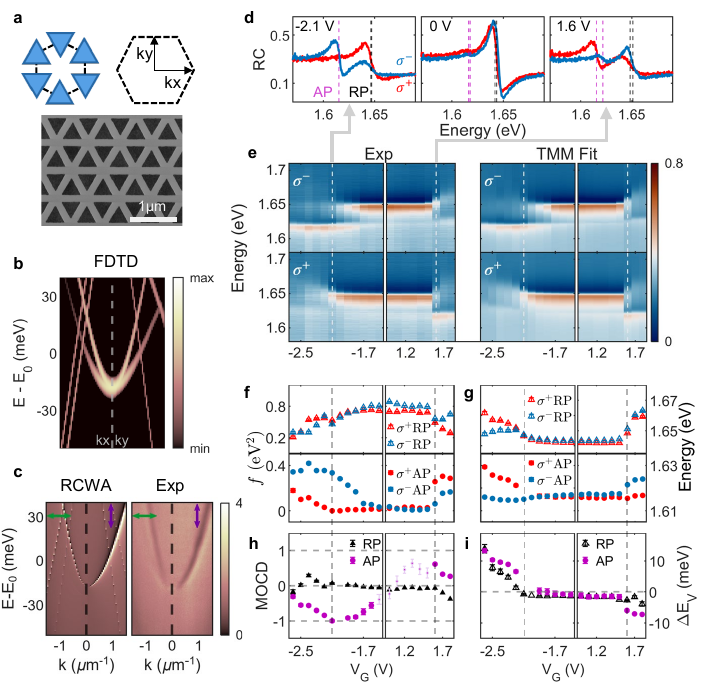}
\caption{\label{f2} \textbf{Experimental characterization of bare PhC and gated-\MoSe ML.} 
\textbf{a}, Schematic of the unit cell (top left), Brillouin zone (top right), and scanning electron microscope image (bottom) of the PhC with $C_6$ symmetry. 
\textbf{b}, PhC bands simulated via FDTD along $k_x$ and $k_y$ as marked in (a). 
$E_0$ is the average of the $\sigma^\pm$ exciton energies at $V_{\mathrm{G}}=0$ V. 
\textbf{c}, Simulated (left) and measured (right) linearly polarized RC spectra of the bare PhC device at $\phi_k=45^\circ$ under two orthogonal incident linear polarizations, indicated by the green and purple arrows. 
\textbf{d}, Circularly polarized RC spectra at representative gate voltages for hole doping ($V_G=-2.1$~V), near neutrality ($V_G=0$~V), and electron doping ($V_G=1.6$~V). Red/blue denote $\sigma^+$/$\sigma^-$; black/magenta dashed lines denote RP/AP. 
\textbf{e}, Circularly polarized RC spectra from the \MoSe ML on the non-PhC region as a function of gate voltage $V_{\mathrm{G}}$. Top and bottom compare $\sigma^-$/$\sigma^+$. Left and right compare experimental data and transfer-matrix-method (TMM) fits. 
\textbf{f-g}, Gate voltage dependence of oscillator strengths (f) and resonance energies (g) of $\sigma^+$ (red) and $\sigma^-$ (blue) polarized RP (triangles, top) and AP (circles, bottom) obtained from (e). 
\textbf{h-i}, Corresponding gate voltage dependence of MOCDs (h) and valley splittings $\Delta E_V$ of RP (black) and AP (purple). Light-markers indicate MOCD data of large uncertainty due to negligible AP oscillator strength.}
\end{figure}

\begin{figure}[th]
\centering
\includegraphics[width=1\linewidth]{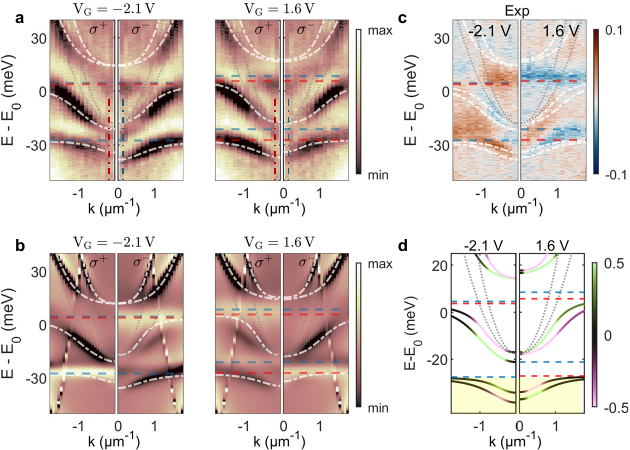}
\caption{\label{f3} \textbf{Topological Fermi-polaron polariton: band structure, circular dichroism and Berry curvature.} 
\textbf{a-b}, Measured (a) and simulated (b) circularly polarized momentum-resolved derivative RC measured from the \MoSe-PhC device for $\sigma^+$ ($k<$0) and $\sigma^-$ ($k>$0) polarization, at $V_{\mathrm{G}}=-2.1$~V (hole doping, left) and $V_{\mathrm{G}}=1.6$~V (electron doping, right), compared with calculated polariton dispersions with appreciable optical visibility in the corresponding detection channel (white dashed lines), showing gap openings at $\Gamma$ between the $\sigma^\pm$ polarizations.  
\textbf{c}, Circular dichroism map of polariton bands from the data shown in (a), for $V_{\mathrm{G}}=-2.1$~V (left) and $1.6$~V (right). 
\textbf{d}, Calculated Berry curvature (color scale) of polariton bands at $V_{\mathrm{G}}=-2.1$~V (left) and $1.6$~V (right). 
In (a-d), $E_0$ is the average of the $\sigma^\pm$ exciton energies at $V_{\mathrm{G}}=0$ V. The blue/red horizontal lines mark uncoupled $\sigma^-$/$\sigma^+$ Fermi polarons; the dotted gray lines mark the bare PhC dispersions.
}
\end{figure}

\begin{figure}[h]
\centering
\includegraphics[width=1\linewidth]{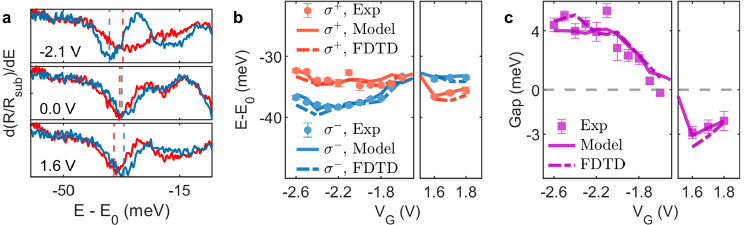}
\caption{\label{f4} \textbf{Electrical control of gap size in polaron-polaritons.} 
\textbf{a}, Line cuts of derivative RC near $k \approx 0$, taken along the dashed vertical lines in Fig.~\ref{f3}\textbf{a}, at three representative voltages. Red/blue represent $\sigma^+$/$\sigma^-$. Vertical lines mark polariton resonances calculated with the Hamiltonian model.
\textbf{b}, $\sigma^+$/$\sigma^-$ (red/blue) lower polariton energies at $k \approx 0$ versus the gate voltage $V_{\mathrm{G}}$. 
\textbf{c}, TRS breaking-induced energy gap between $\sigma^+$ and $\sigma^-$-polarized lower polaritons versus $V_{\mathrm{G}}$. In (b,c), dots are experimental data, solid lines are predictions of Hamiltonian model, and dashed lines are FDTD simulation results. The error bars represent the fitting uncertainties of the resonance centers. 
}
\end{figure}

\clearpage

\end{document}